\documentclass[]{article}

\usepackage{amsmath}
\usepackage{amssymb}
\usepackage{amsthm}
\usepackage{mathtools}
\usepackage{bbm}
\usepackage{algorithm}
\usepackage{algpseudocode} 
\usepackage{verbatim}
\usepackage[margin=0.95in]{geometry}
\usepackage{thmtools}

\usepackage{pdfpages}
\usepackage{graphicx}
\usepackage{wrapfig}
\usepackage{listings}
\usepackage{todonotes}
\usepackage{subcaption}
\usepackage{parskip}
\usepackage{xspace}
\usepackage{tikz}
\usetikzlibrary{shapes.geometric}
\usepackage{wrapfig}
\usetikzlibrary{arrows.meta,positioning}
\usetikzlibrary{decorations.pathreplacing}

\usepackage{hyperref}
\usepackage[capitalise,nameinlink]{cleveref}

\renewenvironment{proof}{{\par\noindent{\bf Proof:}}}{\hfill $\Box$ \\}  
\theoremstyle{definition}
\newtheorem{theorem}{Theorem}[section]
\newtheorem{definition}[theorem]{Definition}
\newtheorem{corollary}[theorem]{Corollary}
\newtheorem{lemma}[theorem]{Lemma}
\newtheorem*{lemma*}{Lemma}

\newtheorem{proposition}[theorem]{Proposition}

\newtheorem{observation}[theorem]{Observation}
\newtheorem*{claim*}{Claim}
\newtheorem{example}[theorem]{Example}
\newtheorem*{example*}{Example}
\newtheorem{remark}[theorem]{Remark}
\newtheorem*{remark*}{Remark}

\newcommand{\network}{$(G,u_0,\nu,\tau)$\xspace}

\newcommand{\RR}{\mathbb R}

\newcommand{\R}{\mathbb R}

\DeclareMathOperator{\intr}{int}

\newcommand{\Rnn}{\RR_{\geq 0}}

\newcommand{\sourceTime}{\theta}
\newcommand{\netTime}{\varphi}
\newcommand{\NFOT}{Nash flow over time\xspace}
\newcommand{\NFOTs}{Nash flows over time\xspace}

\newcommand{\lpref}[1]{(\ref{#1})}

\newcommand{\TT}{\Delta}
\newcommand{\TTT}{$\Delta$\xspace}

\newcommand{\length}{$\tau$-length\xspace}
\newcommand{\longest}{$\tau$-longest\xspace}
\newcommand{\sepa}{series-parallel\xspace}
\newcommand{\steadytime}{T_{SS}}
\newcommand{\instance}{(G,u_0,\nu,\tau,B)}
\newcommand{\perturb}{$\eta$-increase\xspace}

\newcommand{\MNTTP}{\textsc{SSCA}\xspace}
\newcommand{\problemname}{Steady State Capacity Augmentation for Nash Flows over Time\xspace}

\usepackage{graphicx} 

\newbool{withcomments}
\setbool{withcomments}{true}
\ifbool{withcomments}{
\newcommand{\lr}[1]{\textcolor{red!80!black}{\em \small [Laura: #1]}}
\newcommand{\nd}[1]{{\color{orange!90!black}\em \small [Niklas: #1]}}
\newcommand{\rs}[1]{{\color{green!80!black}\em \small [Ruben: #1]}}
\newcommand{\pp}[1]{{\color{blue!80!black}\em \small [Paul: #1]}}
}{
\newcommand{\lr}[1]{ }
\newcommand{\nd}[1]{ }
\newcommand{\rs}[1]{ }
\newcommand{\pp}[1]{ }
}

\newcommand\scl{1.1}

\title{Optimizing the Steady-State Travel Time of Nash Flows over Time with Capacity Increases}

\date{}

\author{ 
Niklas Dahlmeier \thanks{RWTH Aachen, {\textsf{dahlmeier@gdm.rwth-aachen.de}}}
\and
Paul Paschmanns  \thanks{ETH Zurich, {\textsf{paul.paschmanns@inf.ethz.ch}}}
\and
Laura Ritter  \thanks{RWTH Aachen, {\textsf{laura.ritter@oms.rwth-aachen.de}}}
\and
Ruben Skorupinski  \thanks{EPFL, {\textsf{ruben.skorupinski@epfl.ch}}}
}

\begin{document}

\maketitle
\begin{abstract}
  
   We study budgeted capacity augmentation for single-source single-sink \NFOTs under constant inflow $u_0$. The goal is to increase capacities on arcs subject to a global budget constraint in order to minimize the steady-state travel time in the network. We show that this problem is weakly NP-hard on series-parallel networks and provide an exact pseudopolynomial algorithm as well as a fully polynomial algorithm that attains the optimum travel time for budget $B$ while spending at most $B + \varepsilon\cdot\max\{B,u_0\}$. On general networks the problem becomes strongly NP-hard.
   Moreover,  unless $\text{P} = \text{NP}$, for every fixed $\varepsilon \in (0,1)$, no polynomial-time algorithm can achieve the optimal steady-state travel time while exceeding the budget by at most a factor of $n^{1-\varepsilon}$.
   Lastly, we give structural properties about steady-state travel time under varying arc capacities. We identify capacity thresholds determined by subsets of $s$-$t$-cuts and show that the travel time is constant in the interior of the cells resulting from the corresponding hyperplane arrangement. 
\end{abstract}
\newpage

\section{Introduction}
Congestion is a common feature of many real-world network problems, inherent in traffic and communication networks. Here, selfish users choose a path to traverse a network depending on link lengths and congestion induced by other users. This phenomenon can be modeled with a game theoretic variant of flows over time, in which infinitesimally small selfish agents traverse a network according to specified free flow arc traversal times as well as arc capacities. If the capacities in the network are exceeded, queues start to build up, adding to users' travel time. Assuming omniscient users, this leads to the notion of a Nash flow over time, a strategy profile in which all users are traveling on shortest paths.
Crucial structural results on \NFOTs by Koch and Skutella \cite{koch2011nash} (see also \cite{koch2012routing}) have led to many properties of \NFOTs being established for single-source single-sink networks. Key among those are existence \cite{cominetti2015dynamic}, uniqueness of equilibrium travel times \cite{cominetti2015dynamic,olver2022continuity}, continuity \cite{olver2022continuity}, a bound on the price of anarchy in special cases \cite{correa2019price}, and long-term behavior \cite{cominetti2022long,olver2022continuity}: following transient phases, a \NFOT settles into a so-called \emph{steady state} in finite time. In the steady state, queues grow at a constant rate or remain constant \cite{cominetti2022long, olver2022continuity}. A comprehensive overview of Nash flows 

\subsection{Steady State Capacity Augmentation for Nash Flows over Time} 
One key property of \NFOTs is that the steady-state travel time depends non-trivially on the network congestion building up during the transient phases, which in turn depends on the arc capacities. It is therefore natural to ask how the steady-state travel time can be optimized via capacity modifications. We consider the problem of distributing a budget for capacity increases among arcs to optimize the steady-state travel time, referred to as the \emph{\problemname (\MNTTP)}.
Previous work on long-term behavior of \NFOTs has split into two settings, based on the relation between network inflow rate $u_0$ and minimum-cut capacity $\nu_C$. If $u_0> \nu_C$, the network cannot support the entire flow and the steady-state travel time grows linearly \cite{olver2022continuity}. However, minimizing its growth rate reduces to maximizing $\nu_C$ with the budget constraint, a problem solvable in polynomial time. The more interesting regime occurs when $\nu_C$ surpasses $u_0$. Here, the steady-state travel time remains constant over time: we denote it by \TTT. Cominetti, Correa and Olver \cite{cominetti2022long} showed that \TTT can be computed as the solution to a minimum cost flow linear program, if the value of the sink potential $d_t$ is unique among all optimal dual solutions. Despite this simple description in terms of an LP, optimizing \TTT turns out to be hard, even on simple \sepa networks. A key difficulty is that we optimize a single one of the variables appearing in the dual objective by changing the objective coefficients. However, the characterization via linear programming opens up a geometric perspective on the problem, providing structural insights into the behavior of \TTT within the capacity space. In particular, \TTT remains constant within regions defined by the network topology, so any change in \TTT requires crossing some network-dependent capacity threshold. Thus, \MNTTP has a fundamentally discontinuous nature. 

On general networks, increasing capacities is not necessarily beneficial, an instance of Braess' paradox \cite{braess1968paradoxon}. However, the topology of \sepa networks yields both monotone behavior and a characterization of \TTT in terms of dually active paths. These structural results enable us to develop algorithms for the \sepa case. On general networks that can lack such structure and monotonicity \MNTTP is inapproximable in terms of budget.

\subsection{Overview and Results}
In \cref{sec:preliminaries} we introduce the Vickrey bottleneck model, our setting and the terminology.

\cref{sec:hyperplanes} contains structural results which later enable the development of algorithms for the \sepa case.
Fixing the network topology, transit times, and inflow rate, we identify an explicit arrangement of hyperplanes in the capacity space such that the steady-state travel time is constant within every full-dimensional cell. These hyperplanes are defined by signed capacity balances across source–sink cuts. Thus, changes in the steady-state travel time occur only at specific capacity thresholds.
For series-parallel networks, we establish additional structure: away from the exceptional hyperplanes, \TTT equals the minimum longest free-flow source-sink path length over supports of feasible static flows and is non-increasing under capacity increases.

Lastly, \cref{sec:theproblem} considers the problem of increasing capacities in order to minimize steady-state travel time under a budget constraint. We refer to this problem as \MNTTP. Here, we state an idealized version of the problem.

\begin{description}
    \item[\textbf{Idealized \problemname (\MNTTP):}]\hspace*{0cm}\\ Given $\instance$, consisting of a network $G$ with capacities $\nu$ and free flow travel times $\tau$, an inflow rate $u_0$ and a budget $B$, find a capacity vector $\xi\geq 0$ with $\sum_{e\in E} \xi_e \leq B$ such that the Nash flow instance $(G,u_0, \nu + \xi, \tau)$ has minimum steady-state travel time.
\end{description}

To avoid exceptional hyperplanes, the version we study in \cref{sec:theproblem} considers the steady-state travel time after multiplying the final capacities by an arbitrarily small perturbation.

First, we resolve \MNTTP for the case of series-parallel networks. Weak NP-hardness follows from the classic \textsc{Knapsack} reduction. We subsequently give both a pseudopolynomial algorithm as well as a fully polynomial time algorithm with additive budget augmentation $\varepsilon\cdot\max\{B, u_0\}$ in order to obtain a solution with at most the optimal travel time for the original budget. More precisely, given an instance $\instance$, let \TTT be the minimum achievable steady-state travel time. We now ask by how much the budget $B$ has to be relaxed in order to obtain a steady-state travel time of at most \TTT in polynomial time.
In \cref{sec:general_graphs} we show that the problem generally is strongly NP-hard, even on DAGs. Furthermore, we give an $n^{1-\varepsilon}$ budget inapproximability result in a similar sense as the budget augmentation algorithm above, where $n$ is the number of vertices of the constructed \MNTTP instance. We obtain both of these results with a reduction from \textsc{$k$-Disjoint Paths}. 

\subsection{Related Literature}

 While heuristics for the network design problem with capacity modifications have been studied for a long time \cite{yang1998models}, some more recent work has provided algorithms with provable guarantees: Closest to our work are Bhaskar, Fleischer and Anshelevich \cite{bhaskar2015stackelberg} who optimize two price of anarchy measures of a \NFOT by decreasing arc capacities. They obtain a small constant bound on two price of anarchy versions for a fixed flow volume, however, the required capacity decrease can be large. For the problem of optimizing average travel time in a static budgeted capacity improvement setting, Bhaskar, Ligett, and Schulman \cite{bhaskar2014network} show NP-hardness and an FPTAS for series-parallel networks. Both in terms of the objective and results, this work is closest to ours in the static setting. Gairing, Harks, and Klimm \cite{gairing2017complexity} prove strong NP- and APX-hardness as well as approximation results for continuous network design, where investment and equilibrium routing costs are optimized jointly.
However, while the steady state essentially constitutes a static flow, the problem structure of \MNTTP remains fundamentally different from purely static settings as even the steady-state congestion delays in the Vickrey bottleneck model arise from dynamically generated queues.

One key ingredient for our algorithms in the \sepa case is the characterization of \TTT in terms of static flow supports and the resulting monotonicity of \TTT in the capacities, i.e., the absence of Braess' paradox.
In static routing games, the networks immune to Braess' paradox are exactly those without the Braess network as a topological minor, i.e., \sepa networks \cite{chen2015excluding,milchtaich2006network}. In the dynamic case, the \sepa network topology giving rise to the pulse gadget \cite{cominetti2022long} also exhibits Braess-like behaviour during the transient phases of a \NFOT \cite{macko2013braess}. Recently, Chen, Deng and Wang \cite{chen2026network} fully characterized the network topologies immune to Braess' paradox during the entire flow evolution as a subclass of \sepa networks. The same subclass constitutes the most general class of networks for which the monotonicity conjecture is known to hold \cite{chen2026network}. The monotonicity conjecture \cite{correa2019price} asks whether the time required for a fixed flow mass to reach the sink is non-increasing in the inflow rate. It was recently disproved for general networks \cite{Correa2026Counterexample} but remains open for \sepa networks. Thus, optimizing transient together with steady-state travel times likely requires different techniques not relying on monotonicity properties.

\paragraph*{Acknowledgments}
Some of the results in this paper were obtained as part of Ruben Skorupinski's master's thesis at ETH Zurich, under the supervision of Leon Sering. The author is very grateful to Leon for introducing him to the world of Nash Flows over Time and for the many helpful discussions during the writing of the thesis.

\paragraph*{AI Disclosure}
OpenAI GPT 5.6 Sol gave a first idea for the strong NP-hardness proof using 2-disjoint paths. We extended this construction to yield strong NP-hardness on DAGs and inapproximability. We used OpenAI GPT 5.5 to adapt an example in which $\TT$ was non-monotone when modifying the capacities of two arcs to only require a capacity change on a single arc. Several AI tools were used for final polishing of all figures and for proofreading.

\section{Model and Preliminaries}
\label{sec:preliminaries}
We consider the Vickrey bottleneck model \cite{vickrey1969congestion} consisting of a directed network $G=(V,E)$ with a designated source $s\in V$ and sink $t\in V$. Flow enters the network at $s$ at a constant network inflow rate $u_0$. Moreover, each arc $e\in E$ is equipped with a free flow transit time $\tau_e\geq 0$, a bottleneck capacity $\nu_e>0$ and a point queue at the entrance of the arc in which flow accumulates if the inflow into the arc exceeds $\nu_e$. We assume for convenience that every node $v$ lies on some $s$-$t$-path and that the network does not contain any directed cycles of transit time 0. We will refer to this network with all its parameters as \network. We denote the set of arcs entering and leaving a vertex set $S$ by $\delta^-(S)$ and $\delta^+(S)$ respectively, and $\delta(S) = \delta^+(S)\cup\delta^-(S)$, as well as $\delta_A(B)\coloneqq  A \cap \delta(B)$ for some arc set $A$, omitting set notation for single element sets.  We write $\sourceTime$ for a time measured at the source and $\netTime$ for a snapshot time during the evolution. 

\subsection{Flows over Time}
A flow over time is a family of functions $f \in L_{\text{loc},+}^1(\Rnn)^{E \times \{\pm\}}$, where $f_e^+(\netTime)$ and $f_e^-(\netTime)$ denote the in- and outflow rate for arc $e$ at time $\netTime\in \Rnn$. The cumulative inflow into $e$ up to time $\netTime$ is defined as $F_e^+(\netTime)\coloneqq \int_0^\netTime f_e^+(\netTime')d\netTime'$; the definition for the cumulative outflow $F_e^-$ is analogous.
The flow volume in the queue on $e$ at time $\netTime$ can thus be expressed as $z_e(\netTime) \coloneqq F_e^+(\netTime)-F_e^-(\netTime+\tau_e)$. Queues operate at capacity so that for all $e\in E$ and almost all $\netTime\in \Rnn$ \[f_e^-(\netTime+\tau_e) = \begin{cases}
    \nu_e & \text{ if } z_e(\netTime)>0\\
    \min\{f_e^+(\netTime),\nu_e\}&\text{ otherwise.}
\end{cases}\]
The queueing time a flow particle entering arc $e$ at time $\netTime$ experiences on $e$ is thus $q_e(\netTime)\coloneqq \frac{z_e(\netTime)}{\nu_e}$.
A feasible flow over time needs to satisfy flow conservation, i.e., for almost all $\netTime\in \Rnn$: \[\sum_{e\in \delta^+(v)} f_e^+(\netTime)-\sum_{e\in \delta^-(v)}f_e^-(\netTime)\begin{cases}
    = 0 & \text{if }v\in V\setminus\{s,t\}\\
    =u_0 & \text{if }v=s\\
    \leq 0 & \text{if }v=t
\end{cases}\]

\subsection{Nash Flows over Time}
Starting from an empty network, we define the earliest arrival labels for a flow particle entering the network at time $\sourceTime$ via the Bellman equations \[\ell_v(\sourceTime)\coloneqq \begin{cases}
    \sourceTime & \text{ if } v=s\\
    \min_{e=uv\in \delta^-(v)} \ell_u(\sourceTime)+\tau_e+ q_e(\ell_u(\sourceTime)) & \text{ otherwise.}
\end{cases}\]
The set of earliest arrival labels at time $\sourceTime$ defines the set of active arcs $E'(\sourceTime)$ and resetting arcs $E^*(\sourceTime)$: 
\[E'(\sourceTime)\coloneqq \{e=uv\in E \mid \ell_v(\sourceTime)=\ell_u(\sourceTime)+\tau_e+q_e(\ell_u(\sourceTime))\}\quad E^*(\sourceTime)\coloneqq \{e=uv\in E \mid q_e(\ell_u(\sourceTime))>0\}\]
Intuitively, active arcs lie on a shortest $s$-$v$-path for some $v\in V$ as seen by a particle entering the network at time $\sourceTime$; on resetting arcs such a particle would encounter a queue if it enters the arc at the earliest possible time.
A dynamic equilibrium, or Nash flow over time, is a feasible flow over time in which almost all particles travel on a shortest $s$-$t$-path.
\begin{definition}
    A Nash flow over time is a feasible flow satisfying $f_e^+(\netTime)> 0 \implies \netTime\in \ell_u(\Theta_e)$ for each arc $e=uv\in E$ and almost all times $\netTime\in \Rnn$, where $\Theta_e\coloneqq\{\sourceTime\in \Rnn\mid e\in E'(\sourceTime)\}$ is the set of network entry times for which arc $e$ is active.
\end{definition}
\subsection{Setting}
\subsubsection{Inflow Rate Bounded by Minimum Cut Capacity}
In our single-source, single-sink setting, Nash flows over time exist and are unique in the sense that the earliest arrival labels are unique \cite{cominetti2015dynamic}. 

A key property of \NFOTs is that there exists a finite time $\steadytime$ such that a \emph{steady state} is reached \cite{cominetti2022long,olver2022continuity}. Upon reaching the steady state, the queueing times and thus the earliest arrival labels $\ell_v$ grow linearly at a constant rate.
In particular, by Lemma 6.34 in \cite{koch2012routing} and Theorem 3.3 in \cite{olver2022continuity} the growth rate of the steady-state travel time is $ \max\{0,\tfrac{u_0}{\nu_C}-1\}$, where $\nu_C$ is the capacity of a minimum cut.
If $u_0 > \nu_C$, the main contribution to the steady-state travel time is the positive growth rate. Thus, in this case, optimizing the steady-state travel time 
asymptotically amounts to minimizing the slope of its linear growth. This is equivalent to maximizing $\nu_C$, which can be done by including variables for budget assignment into a max-flow LP. The resulting LP still has polynomial size and can thus be solved efficiently.

Thus, we restrict to the more interesting case $u_0 \leq \nu_C$ for the remainder of this work and optimize the intercept instead of the slope of the affine linear function describing the steady-state travel time. 
We denote the \emph{steady-state travel time} by \TTT. Given a network with data $(G,u_0,\tau)$, we write $\steadytime(\nu)$ for the time to reach steady state and $\TT(\nu)$ for the steady-state travel time in \network. All subsequent results are contingent on the assumption that any capacity vector $\nu$ we consider yields a minimum cut with capacity at least $u_0$.

The steady-state behavior of Nash flows over time can be characterized by the primal and dual linear programs for minimum cost flows \cite{cominetti2022long}; here, $\mathcal{F}_{u_0}$ refers to the set of static $s$-$t$-flows of value $u_0$ and is used as shorthand for the corresponding flow conservation constraints. 

\begin{minipage}[t]{0.4\textwidth}
    \centering
        \begin{align*}
            & \textbf{Primal:} \tag{P} \label{P}\\
            & \min_{f\in \RR^{E}}   \quad && \sum_{e\in E}\tau_e\cdot f_e \\
            & \text{s.t.}  \quad && f \in \mathcal{F}_{u_0} \\
            &              \quad && 0 \leq f_e \leq {\nu_e} \quad 
            \forall e\in E
        \end{align*}
        \end{minipage}
        \hspace{0.5cm}
        \begin{minipage}[t]{0.4\textwidth}
        \centering
        \begin{align*}
            & \textbf{Dual:} \tag{D} \label{D}\\
            & \max_{\substack{d \in \RR^{V} \\ q \in \RR^{E}}}   \quad && u_0d_t- \sum_{e\in E}{\nu_e}\cdot q_e \\
            & \text{s.t.}  \quad && d_s = 0 &\\
            &              \quad && d_v \leq d_u +\tau_e +q_e \quad &&
            \forall e=uv \in E\\
            &              \quad && q_e\geq 0   \quad && 
            \forall e \in E
        \end{align*}
\end{minipage}
\vspace{0.2cm}

The interpretation of the dual in terms of normalized earliest arrival labels is straightforward: having fixed $d_s=0$, the label $d_v$ can be seen as the shortest possible travel time from $s$ to $v$ in the steady state; $q_e$ gives the queueing times in steady state. Our main interest lies in the value of $d_t$ as there always exists an optimal dual solution with $d_t=\TT$ \cite{cominetti2022long}. In particular, \TTT can be obtained from \lpref{D} whenever $d_t$ is unique.

\begin{figure*}[t!]\label{fig:unnec_queue}
    \centering
    \begin{subfigure}[t]{0.45\textwidth}
        \centering
         \begin{tikzpicture}[
            >=Stealth,
            node/.style={
                circle,
                draw,
                inner sep=1.8pt
            },
            arc/.style={
                ->,
                thick,
                line cap=round
            },
            label/.style={
                font=\small
            },
            edge label/.style={
                font=\scriptsize,
                inner sep=1pt,
                fill=white
            }
        ]
        
        \node[node] (s) at (0,2.0) {$s$};
        \node[node] (v) at (1.5,0) {$v$};
        \node[node] (w) at (3.5,2.0) {$w$};
        \node[node] (t) at (5.5,2.0) {$t$};

        \draw[arc]
            (-0.75,2.0) to 
            node[edge label,above,pos=0.30] {$12$}
            (s);
            
        \draw[arc]
            (s) to[out=40,in=150]
            node[edge label,above,pos=0.50] {$(4,1)$}
            node[edge label,below,pos=0.50] {$e_1$}
            (w);
        \draw[arc]
            (s) to[out=-80,in=150]
            node[edge label,above right ,pos=0.45] {$(9,0)$}
            node[edge label,below left,pos=0.45] {$e_2$}
            (v);
        
        \draw[arc]
            (v) to[out=70,in=195]
            node[edge label,below right,pos=0.55] {$e_3$}
            node[edge label,above left,pos=0.55] {$(2,0)$}
            (w);
        \draw[arc]
            (v) to[out=10,in=-105]
            node[edge label,below right,pos=0.50] {$e_4$}
            node[edge label,above left,pos=0.52] {$(12,1)$}
            (w);
        \draw[arc]
            (w) to[out=0,in=180] 
            node[edge label,above=2pt,pos=0.5] {$(12,0)$}
            node[edge label,below=2pt,pos=0.5] {$e_5$}
            (t);

            \end{tikzpicture}
            \caption{Pulse gadget \cite{cominetti2022long} with unnecessary queue on $e_5$; the minimum cut capacity equals the inflow rate.}
            \label{fig:unnecessary queue}
    \end{subfigure}%
    \hfill
    \begin{subfigure}[t]{0.45\textwidth}
        \centering \begin{tikzpicture}[
                        >=Stealth,
                        node/.style={
                            circle,
                            draw,
                            inner sep=1.8pt
                        },
                        arc/.style={
                            ->,
                            thick,
                            line cap=round
                        },
                        label/.style={
                            font=\small
                        },
                        edge label/.style={
                            font=\scriptsize,
                            inner sep=1pt,
                            fill=white
                        }
                    ]
                    
                    \node[node] (s) at (0,0) {$s$};
                    \node[node] (v) at (3,1.5) {$v$};
                    \node[node] (w) at (3,-1.5) {$w$};
                    \node[node] (t) at (6,0) {$t$};

                    \draw[arc]
                        (-0.75,0) to 
                        node[edge label,above,pos=0.5] {$1$}
                        (s);
                        
                    \draw[arc]
                        (s) to
                        node[edge label,below=5pt,pos=0.4] {$(2,3)$}
                        (w);
                    \draw[arc]
                        (s) to
                        node[edge label,below =7pt,pos=0.7] {$(0.6,0)$}
                        (v);
                    \draw[arc]
                        (s) to[out=70,in=160]
                        node[edge label,above=9pt ,pos=0.3] {$(2,1)$}
                        (v);
                     \draw[arc]
                        (s) to
                        node[edge label,above,pos=0.7] {$(2,4)$}
                        (t);
                    \draw[arc]
                        (v) to
                        node[edge label, left,pos=0.7] {$(\nu_{vw},0)$}
                        (w);
                    \draw[arc]
                        (v) to
                        node[edge label,above=5pt,pos=0.5] {$(2,3)$}
                        (t);
                    \draw[arc]
                        (w) to
                        node[edge label,above=7pt,pos=0.3] {$(0.6,0)$}
                        (t);
                    \draw[arc]
                        (w) to[out=-20,in=-110]
                        node[edge label,below=9pt,pos=0.7] {$(2,1)$}
                        (t);
                        \end{tikzpicture}
                        \caption{Network exhibiting Braess' paradox and behavior that is non-monotonic in $\nu_{vw}$.}
                        \label{fig:braess}
    \end{subfigure}
    \caption{Examples for complex behavior of \NFOTs with arc data $(\nu,\tau)$.}
\end{figure*}
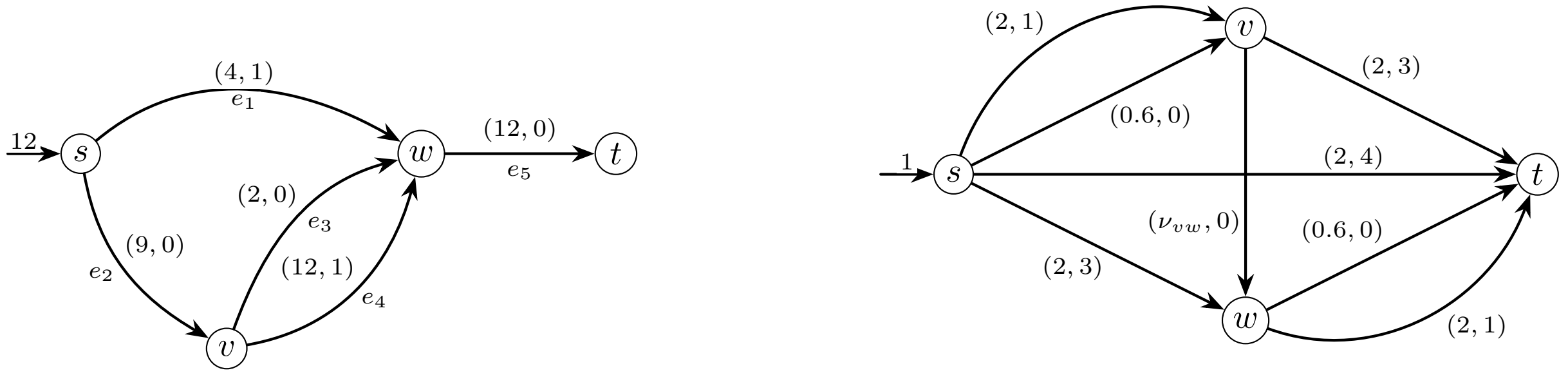

\begin{example}\label{ex:unnec_queue}
    Consider the network depicted in \cref{fig:unnecessary queue}, consisting of a pulse gadget \cite{cominetti2022long} composed with a single arc $e_5$. The pulse gadget has the property that for a short time interval during the evolution of the Nash flow over time the total outflow rate out of arcs $e_1,e_3$ and $e_4$ exceeds $u_0$. In the present network, this pulse then causes a queue to build up on arc $e_5$ which can never deplete, as after the pulse the inflow into $e_5$ is equal to its capacity. The queue on $e_5$ and the additional queueing time it induces in the steady state is in some sense ``unnecessary'' as increasing the capacity of arc $e_5$ by any $\varepsilon>0$ leads to the depletion of the queue and thus its absence in the steady state. The $d_t$ value for this network is not unique as increasing both $q_{e_5}$ and $d_t$ by the same value results in a different optimal dual solution.
\end{example}

\subsubsection{Series-Parallel Networks}
Series-parallel networks are a fundamental class of networks in flow theory because their recursive structure facilitates analysis and algorithm design.

\begin{definition}
    A directed \sepa network is a network with a source $s$ and a sink $t$ which either consists of a single arc $st$ or can be obtained by a series or parallel composition from two \sepa networks. Given two \sepa networks with sources $s_1,s_2$ and sinks $t_1,t_2$, a series composition merges $t_1$ and $s_2$ to a single node, whereas a parallel composition merges $s_1$ with $s_2$ and $t_1$ with $t_2$. Note that parallel arcs between two nodes are allowed.
\end{definition}

Despite the simple structure of \sepa networks, \NFOTs can exhibit complex behavior even on this class of networks, as in \cref{ex:unnec_queue}.

\section{Structural Results}
\label{sec:hyperplanes}
\subsection{General Networks}
The goal of this section is to analyze the steady state travel time $\TT$ in a given network under changes of its capacities $\nu$. First, observe that uniqueness of the $d_t$ value in an optimal solution $(d,q)$ to \lpref{D} is determined by the set of active and resetting arcs in the steady state. 

\begin{proposition}\label{pr:non-uniqueness}
    There exist distinct $d_t$-values among the set of optimal solutions to the LP \lpref{D} for a network \network if and only if there exists an optimal solution $(d,q)$ with active and resetting arcs $ E',E^*$, and an $s$-$t$-cut $S\subseteq V$ such that $\delta_{E'}(S)\subseteq E^*$.
\end{proposition}

Intuitively, if all flow must travel via arcs which already have a queue, as these are the only active arcs across the cut, then there exists a small modification of the queueing times such that no inactive arc becomes active and no queue is removed completely. This argument can be made formal, however, we instead present the following concise LP-based proof.

\begin{proof}
    Let $F$ be the face of all optimal solutions to \lpref{D}, and let $A,B\subseteq E$ index the constraints tight throughout $F$. The equalities
    \begin{align*}
        d_s &= 0,\\
        d_v &= d_u+\tau_e+q_e &&\forall e\in A,\\
        q_e &= 0 &&\forall e\in B
    \end{align*}
    define the affine hull of $F$. Eliminating $q_e$ for $e\in A\setminus B$ shows that these equalities fix $d_t$ if and only if $A\cap B$ contains an undirected $s$-$t$-path. At an optimal solution $(d,q)$ in the relative interior of $F$, we have $A\cap B=E'\setminus E^*$. Thus, if $d_t$ is not constant on $F$, there is no such path, and hence there exists an $s$-$t$-cut $S$ with $\delta_{E'}(S)\subseteq E^*$. Conversely, $A\cap B\subseteq E'\setminus E^*$ at every optimal solution. Therefore, the existence of such a cut at any optimal solution rules out an undirected $s$-$t$-path in $A\cap B$, so $d_t$ is not constant on $F$.
\end{proof}

\begin{corollary}\label{cor:active_cut}
    If there exist optimal solutions to the LP \lpref{D} with distinct $d_t$-values in a network \network, then there exists an $s$-$t$-cut $S$ and an optimal steady state solution with active arcs $E'$ such that $\nu(\delta^+_{E'}(S))-\nu(\delta^-_{E'}(S))=u_0$.
\end{corollary}
\begin{proof}
    By \cref{pr:non-uniqueness}, there exists an $s$-$t$-cut $S$ and an optimal dual solution with active arcs $E'$ and resetting arcs $E^*$ such that $\delta_{E'}(S)\subseteq E^*$.
    Consider an optimal primal solution $f$. By complementary slackness, $f_e = \nu_e$ for all $e\in E^*$. Moreover, as $f$ is a flow of value $u_0$ and since $f_e=0$ for all $e \notin E'$, we have $u_0 = f(\delta_{E'}^+(S))- f(\delta_{E'}^-(S)) = \nu(\delta_{E'}^+(S))-\nu(\delta_{E'}^-(S))$.
\end{proof}

The converse of \cref{cor:active_cut} does not hold. Consider the graph $G=(\{s,v,w,t\},\{sv,vw,wt\}$), that is, a path with three arcs. Given the inflow rate $u_0$ we choose capacities $(\nu_{sv},\nu_{vw},\nu_{wt}) = (2u_0,3u_0,2u_0)$ and consider the $s$-$t$-cut $S = \{s,w\}$. Then, $\nu(\delta_{E'}^+(S))-\nu(\delta_{E'}^-(S)) = 2u_0 + 2u_0 -3u_0 = u_0$. However, the unique minimum cost flow $f$ of value $u_0$ does not saturate any arc so any dual optimal steady state solution yields  $E' = E$ and $E^* = \emptyset$. By \cref{pr:non-uniqueness}, $d_t$ is unique.
\\

By \Cref{cor:active_cut}, the value $d_t$ can only be non-unique if the capacity vector lies on one of finitely many hyperplanes defined by (subsets of) $s$-$t$ cuts. We will henceforth consider the geometry of these hyperplanes.

\begin{corollary}
    The set of vectors $\nu \in \RR^E$ for which the $d_t$-value of (\ref{D}) is not unique on \network forms a null-set. More specifically, these vectors are a subset of the following collection of hyperplanes:
    \begin{align*}
        \mathbb{H}_G \coloneqq \bigcup_{\substack{s\text{-}t\text{-cut } S\subseteq V \\ A \subseteq \delta^+(S)\\ B \subseteq \delta^-(S)}}H_{A,B}
        \coloneqq \bigcup_{\substack{s\text{-}t\text{-cut } S\subseteq V \\ A \subseteq \delta^+(S)\\ B \subseteq \delta^-(S)}}\{\nu \in \RR^E: \nu(A) - \nu(B) = u_0\}   \end{align*}
\end{corollary}

In the following we refer to $\mathbb{H}_G$ as a \emph{hyperplane arrangement} and to the resulting $|E|$-dimensional (closed) polyhedra as cells of $\mathbb H_G$. Finally, we call two cells $C$ and $C'$ adjacent if there exists a hyperplane $H \in \mathbb H_G$ in the arrangement such that their intersection with the hyperplane $C\cap H = C' \cap H$ is an $(|E|-1)$-dimensional face of both cells.

\par

The above results already imply that $\Delta$ attains only finitely many values outside $\mathbb{H}_G$ since the $d_t$-values are restricted to the lengths of residual network paths. The next theorem makes this more concrete, proving that in every connected component of $\R^E \setminus \mathbb{H}_G$, i.e., in the interior of every cell, the $d_t$-value of an optimal dual solution $(d,q)$ stays constant. For $\nu\notin \mathbb H_G$ we denote by $d_t(\nu)$ the unique $d_t$ value in an optimal dual solution for \network.

\begin{theorem}\label{thm:d_t_constant_in_region}
    Let $\nu,\bar\nu \notin \mathbb H_G$ for $(G,u_0,\tau)$. If $d_t(\nu)\neq d_t(\bar \nu)$, then $\nu$ and $\bar\nu$ are separated by a hyperplane.
\end{theorem}

\begin{figure}[]
    \begin{subfigure}[t]{0.45\textwidth}
        \centering
        \includegraphics[width=\linewidth]{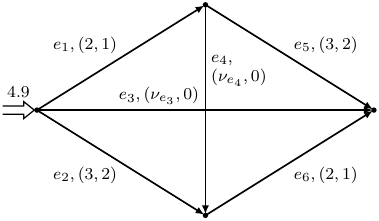}
        \caption{The Braess network with data $(\nu_e,\tau_e)$.}
        \label{fig:monotonicity_counterex_graph}
    \end{subfigure}
    \hfill
    \begin{subfigure}[t]{0.5\textwidth}
        \centering
        \includegraphics[width=\linewidth]{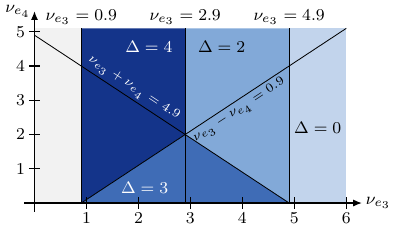}
        \caption{The value of \TTT for varying $\nu_{e_3}$ and $\nu_{e_4}$ in the network to the left. In the left gray region, $u_0>\nu_C$.}
        \label{fig:monotonicity_counterex_plot}
    \end{subfigure}
    
    \vspace{0.5cm}
    
    \begin{subfigure}{1\textwidth}
        \centering
        \includegraphics[width=\linewidth]{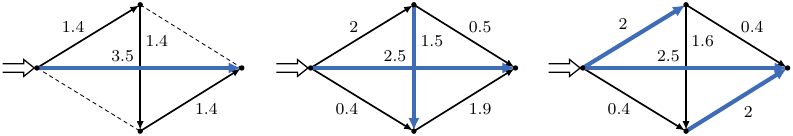}
        \caption{Optimal steady state flows in the network in \ref{fig:monotonicity_counterex_graph} for varying capacities. From left to right the capacities $(\nu_{e_3},\nu_{e_4})$ are $(3.5,2), (2.5,1.5)$ and $(2.5,2)$ leading to $\Delta$-values of 2, 3 and 4. Dashed arcs are non-active, black arcs are active but non-resetting and blue arcs are resetting.}
        \label{fig:monotonicity_counterex_thin_flows}
    \end{subfigure}
    
    \caption{Local behavior of $\Delta$ as a function of the capacities $\nu$ as described in \Cref{thm:d_t_constant_in_region}. On this specific non-series-parallel network (the Braess network) $\Delta$ is not monotone in $\nu$.}
\end{figure}

\begin{proof}
    Let $F_\nu, F_{\Bar{\nu}}$ denote the faces of the feasibility polyhedron $P$ of the dual LP \lpref{D} that are optimal for $\nu$ and $\Bar{\nu}$ respectively. Since $\nu, \Bar{\nu} \notin \mathbb{H}_G$, all vertices within either of the two faces share the same $d_t$-value. In particular, since
    $d_t(\nu)\neq d_t(\bar \nu)$, the two faces are disjoint. Now, consider the line segment $L = \lambda \nu + (1-\lambda)\Bar{\nu}$, for $\lambda \in [0,1]$ and let $\nu_\lambda $ denote a point on $L$. As $\lambda$ varies from $0$ to 1 we exhibit a sequence $0 < \lambda_1< \dots < \lambda_p<1$ as well as vertices $x_1,...,x_{p+1}$ of the polyhedron $P$ such that for every $\lambda_i$ both vertices $x_i$ and $x_{i+1}$ are optimal solutions to \lpref{D}. Since $d_t(\nu) \neq d_t(\Bar{\nu})$ there exists an index $1 \leq i \leq p$ where $d_t(x_{i}) \neq d_t(x_{i+1})$. Since $x_i$ and $x_{i+1}$ are optimal for $\nu_{\lambda_i}$ we deduce from \Cref{cor:active_cut} that $\nu_{\lambda_i} \in \mathbb{H}_G$. Thus, $\nu$ and $\Bar{\nu}$ must lie in two distinct cells of the hyperplane arrangement $\mathbb{H}_G$ since otherwise the entire line segment $L$ would also lie in the shared cell, i.e., $L \cap \mathbb{H}_G = \emptyset$, a contradiction.
\end{proof}
The converse of \cref{thm:d_t_constant_in_region} does not hold: in \cref{fig:monotonicity_counterex_plot} the cells separated by the hyperplane at $\nu_{e_3}=2.9$ only have differing \TTT-values for the region of capacity space where $\nu_{e_4}\geq 2$. 

The following result shows that while we cannot easily determine $\TT(\nu)$ when $\nu \in H \subseteq \mathbb{H}_G$ as discussed in \Cref{ex:unnec_queue}, we can still bound $\TT$ from above and below by the values on two cells adjacent to $H$ at $\nu$. This is especially important since it allows us to perturb $\nu$ slightly to obtain $(1+\eta)\nu \notin \mathbb{H}_G$ for some $\eta > 0$ such that $\Delta((1+\eta)\nu) \leq \Delta(\nu)$. We will refer to such a perturbation as an \emph{\perturb}. 

\begin{restatable}{lemma}{multiplicativeeps}
\label{prop:multiplicative_eps}
Consider \network with $u_0 <\nu_C$. There exists $\eta>0$ small enough such that the sink potential $d_t$ under capacities $(1+\eta)\nu$, respectively $(1-\eta)\nu$, is unique and equal to the minimal, respectively maximal possible, sink potential $d_t$ under $\nu$.

\end{restatable}
The proof is deferred to the appendix.
Intuitively, if a face of positive dimension is optimal for the dual \lpref{D}, perturbing the objective of \lpref{D} slightly, some lower dimensional face of the solution polyhedron remains optimal. The specific perturbation is chosen to guarantee that $d_t$ is unique. An additive perturbation may not have the desired effect if the number of arcs entering and leaving the relevant cut is the same.

\subsection{Series-Parallel Networks}
 In this section we establish two desirable properties of \TTT on \sepa networks: monotonicity of \TTT in $\nu$ and a characterization of \TTT as the length of the longest active path in the network in terms of free-flow transit times. These results are crucial for the algorithms we present in \cref{sec:sepa}.
The advantage of \sepa networks, formalized in \cref{pr:empty_in}, is that in some sense flow only moves ``forwards'' through the network. 

\begin{lemma} \label{pr:empty_in}
    Let $G=(V,E)$ be a series-parallel network and let $A, B \subseteq E$. Assume there exists an $s$-$t$-cut $S \subseteq V$ such that $\delta^+_A(S) \subseteq B$. Then, there exists a second $s$-$t$-cut $S'$ which still satisfies  $\delta^+_A(S') \subseteq B$ but also $\delta^-(S') = \emptyset$.
\end{lemma}
\begin{proof}
    We show the claim by induction over the size of $G$.
    It holds for the graph consisting of just two vertices $s$ and $t$ and a single arc connecting the two as there is only one $s$-$t$-cut and no backwards arcs crossing it. Now, consider a series-parallel graph $G=(V,E)$ and define sets $A_i = A \cap E_i$ and $B_i = B \cap E_i$. Let $S\subseteq V$ be an $s$-$t$-cut in $G$ satisfying $\delta^+_{A}(S)\subseteq B$.    
    Assume that $G$ is the series composition of some $G_1$ and $G_2$ with sources $s_1,s_2$ and sinks $t_1,t_2$. Since $S$ is an $s$-$t$-cut, $S_i \coloneqq S\cap V_i$ is an $s_i$-$t_i$-cut for one $i \in \{1,2\}$ satisfying $\delta^+_{A_i}(S_i) \subseteq B_i$. Let $S'_i$ be the extended cut produced by the induction hypothesis on $G_i$. This gives the desired $s$-$t$-cut $S'$ in the full graph $G$ by setting $S' = S_1'$ or $S' = V_1 \cup S_2'$.
     If $G$ is the parallel composition of $G_1$ and $G_2$, the desired cut is the union of the two extended cuts obtained by applying the induction hypothesis on the cuts $S_i \coloneqq S\cap V_i$ as these are both $s_i$-$t_i$-cuts in $G_i$.
\end{proof}

\vspace*{-4mm}
This forward-flowing property directly leads to two corollaries simplifying the hyperplane arrangements and the set of values $\Delta$ can attain on \sepa networks.

\begin{corollary}\label{prop:uniqueness_char_for_SP-graphs}
    Let \network be a series-parallel network. There exist distinct $d_t$-values among the set of optimal solutions to the LP \lpref{D} if and only if there exists an optimal steady state solution $(d,q)$ with active arcs $E'$ and resetting arcs $E^*$ as well as an $s$-$t$-cut $S\subseteq V$ satisfying $\delta_{E'}^+(S) \subseteq E^*$.
\end{corollary}

\begin{proof}
   Set $A= E', B=E^*$ in \cref{pr:empty_in} and combine with \cref{pr:non-uniqueness}.
\end{proof}

\vspace*{-4mm}
In the following, we consider \network such that $\nu\notin \mathbb H_G$ throughout.
The following observation, while obvious from the shortest path labels in \lpref{D}, is immensely useful. Here, the \length of a path is measured in terms of the free-flow transit times $\tau$.
\begin{observation}\label{obs:longest_path_lower_bound}
    In any network, \TTT is at least the \length of the longest active $s$-$t$-path.
\end{observation}
Crucially, on \sepa networks, this is not only a lower bound, but an equality.
\begin{corollary}\label{cor:sepa_longest_path}
    In a \sepa network \network with $\nu\not\in\mathbb{H}_G$, \TTT is equal to the \length of the \longest active $s$-$t$-path.
\end{corollary}
\begin{proof}
     The lower bound is given by \cref{obs:longest_path_lower_bound}. If \TTT was strictly larger, there would exist a cut such that $\delta_{E'}^+(S) \subseteq {E^*}$. As the network is \sepa, this contradicts $\nu\not\in\mathbb{H}_G$ by \cref{prop:uniqueness_char_for_SP-graphs}.
\end{proof}

\vspace*{-4mm}
This equality does not hold in general networks, as the third network in \cref{fig:monotonicity_counterex_thin_flows} shows. The \longest path in the network has length 3, but the steady state travel time is 4.

\begin{lemma}\label{lm:support_travel_time}
    In a \sepa network with $\nu\notin \mathbb H_G$, let $L_f$ be the maximum \length of any $s$-$t$-path in the support of a feasible flow $f$ of value $u_0$. Then $\TT = \min_{f\in \mathcal F_{u_0}, f\in [0,\nu]}L_f$.
\end{lemma}
\begin{proof}
    As the steady state flow $f^*$ is feasible and its support only contains active arcs, \cref{obs:longest_path_lower_bound} gives $\TT \geq \min_{f\in \mathcal F_{u_0}, f\in [0,\nu]}L_f$.
    Now consider any feasible flow $f$ and assume that $\TT > L_f$. Then all $s$-$t$-paths in $\text{supp} (f)$ contain a queueing arc. Otherwise, a non-resetting $s$-$t$-path $P$ implies $\TT = d_t\leq\sum_{e\in P}\tau_e \leq L_f$, a contradiction. Thus, there is an $s$-$t$-cut $S$ which satisfies $\delta^+_{\text{supp} (f)}(S) \subseteq E^*$. By \cref{pr:empty_in} we may assume without loss of generality that $\delta^-(S) = \emptyset$. As $S$ is an $s$-$t$-cut, the entire flow $f$ passes through $S$, so $\nu(\delta^+_{\text{supp} (f)}(S)) \geq u_0$, but all $S$-leaving arcs in $\text{supp} (f)$ are queueing and hence saturated by $f*$, implying $\nu(\delta^+_{\text{supp} (f)}(S)) \leq u_0$ so $\nu \in \mathbb H_G$, a contradiction.
\end{proof}

\vspace*{-4mm}
Finally, we obtain monotonicity of $\Delta$ as a function of $\nu$ on \sepa networks. 

\begin{lemma}\label{prop:monotonicty_in_s-p_networks}
    Let $\nu_1 , \nu_2 \notin \mathbb{H}_G$ for a \sepa network $(G,u_0,\tau)$.
    Then, $\nu_1 \leq \nu_2$ implies $\TT(\nu_1) \geq \TT(\nu_2)$, i.e. \TTT is monotonically non-increasing in $\nu$.
\end{lemma}

\begin{proof}
This follows by \cref{lm:support_travel_time} as any flow feasible under $\nu_1$ remains feasible under $\nu_2$.
\end{proof}

\vspace*{-4mm}
Again, this statement does not hold for general networks. In the network in \cref{fig:braess} it suffices to change the capacity $\nu_{vw}$ of the single arc $vw$ to observe non-monotone \TTT: for $\nu_{vw}=0.1$ the steady state travel time is 3, for $\nu_{vw}=0.3$ it is 4 and for $\nu_{vw}=1.1$ it is 2.

\section{Optimizing Steady-State Travel Time}
\label{sec:theproblem}

	In this section we consider the problem of minimizing the steady-state travel time \TTT when we are allowed to increase capacities on arcs. More precisely, we are given a budget $B$ with which we can increase capacities on arcs by $\xi\in \RR^E_{\geq 0}$ satisfying $\xi(E) \coloneqq \sum_{e \in E}\xi_e \leq B$.

	\begin{definition}
		An instance $\mathcal{I}=\instance$ of the \textbf{\problemname (\MNTTP)} consists of a network $G$ with designated source and sink, an inflow rate $u_0$, capacities $\nu$, travel times $\tau$ and a budget $B>0$.
        The task is to compute capacity increments $\xi \in \RR_{\geq 0}^E$ which obey the budget constraint $\xi(E) \leq B$ such that for every sufficiently small $\eta>0$ the steady-state travel time $\TT$ in the network $G$ with capacities $(\nu+\xi) \cdot (1+\eta)$, travel times $\tau$, and inflow $u_0$ is minimized. 

	\end{definition} 
    \begin{remark}
        A simple exponential time algorithm is given by enumerating the vertices of the dual feasibility polyhedron $P$, and checking if the capacities $\nu$ can be increased to reach each vertex of the polyhedron. This can easily be done using a linear program, and the number of vertices of $P$ can be upper bounded by $4^{|E|}$ yielding an exact algorithm for \MNTTP running in time $O(4^{|E|}\cdot\text{poly}(|E|))$.
    \end{remark}
    \begin{remark}
        The motivation for the scaling by $1 + \eta$ in the definition of \MNTTP is due to \Cref{prop:multiplicative_eps}: a sufficiently small \perturb ensures that $\Delta$ equals the unique $d_t$ from \lpref{D}. In the following, whenever we invoke a statement requiring $\nu\notin \mathbb H_G$, we apply it to the capacities plus an \perturb.
    \end{remark}

    We consider polynomial-time algorithms obtaining the optimal \TTT while violating the budget by some small additive term. Indeed, if $B\geq u_0$, the resulting algorithm constitutes an FPTAS with the approximation in the budget.

    \begin{definition}[Budget Augmentation Algorithm]
    \label{def:budget-approximation}
        Let $\TT_{\mathcal{I}}$ denote the optimum steady-state travel time for the \MNTTP instance $\mathcal{I}=\instance$. An $\alpha$-budget augmentation algorithm computes in polynomial time capacity increments $\xi \in \RR_{\geq 0}^E$ with $\xi(E) \leq B+ \alpha\cdot \max\{B,u_0\}$ that result in a steady-state travel time of at most $\TT_{\mathcal{I}}$.
    \end{definition}

   Using a standard reduction from \textsc{\textsc{Knapsack}}, we show that the NTP is weakly NP-hard even when restricted to \sepa networks. 
    Subsequently, we give both a dynamic program with pseudopolynomial runtime solving the problem exactly and a fully polynomial time algorithm with additive budget augmentation $\varepsilon\cdot\max\{B,u_0\}$ for any fixed $\varepsilon > 0$. 
    Lastly, we obtain strong NP-hardness and $n^{1-\varepsilon}$-inapproximability even for acyclic non \sepa networks using a reduction from \textsc{$k$-Disjoint Paths}. Here, $n$ is the number of nodes in $G$.

\subsection{\MNTTP on Series-Parallel Networks}
\label{sec:sepa}
\subsubsection{Weak NP-Hardness of \MNTTP on Series-Parallel Networks}

In this section, we show that the \MNTTP is weakly NP-hard on \sepa networks by a reduction from \textsc{Knapsack}.
	
	\begin{theorem}
		The \MNTTP on \sepa networks is weakly NP-hard.
	\end{theorem}
	\begin{proof}
        Given a \textsc{Knapsack} instance consisting of a set of items $[n]$ of positive weight $w_i $ and value $v_i$ for $i\in [n]$ and a capacity constraint $W$, we define an instance of \MNTTP as follows. We set $B \coloneq W$, $u_0 \coloneq \max_{i\in [n]} w_i + 1$ and define the graph $G = (V, E)$ with $V = \{1, ..., n+1\}$ and $E = \bigcup_{i = 1}^{n} \{e_i, e_i'\}$ where $e_i$ and $e_i'$ are parallel arcs directed from node $i$ to $i+1$ with $(\nu_{e_i}, \tau_{e_i}) = (u_0, M)$ and $(\nu_{e'_i}, \tau_{e'_i}) = (u_0 - w_i, M-v_i)$ for $M\geq \max_i v_i$. 

        We show that a solution for the constructed \MNTTP instance with $\TT= nM-K$ yields a feasible solution of value $K$ to the \textsc{Knapsack} instance.        
		In the \MNTTP instance without any capacity increase we have $\TT=n  M$ as $\nu_{e_i'}<u_0$ for all $i$, so $e_i'$ grows a queue until $e_i$ becomes active. We refer to increasing the capacity of $e_i'$ by $\xi_{e_i'}\geq w_i$ to $u_0$ as \emph{buying} $i$. Increasing the capacity on $e_i'$ without buying $i$ has no effect on \TTT. 
		Buying $i$ implies that the entire flow can travel along arc $e_i'$, which decreases the travel time from node $i$ to $i+1$ by $v_i$. Summing over all $I \subseteq \{1,\dots, n\}$ bought in the \MNTTP solution, we obtain $\TT=n M - \sum_{i \in I} v_i = n M - K$.		 
		But clearly $I$ is a feasible solution of value $K = \sum_{i \in I} v_i$ to the \textsc{Knapsack} instance, as we have $\sum_{i \in I} w_i \leq\xi(E) \leq B = W$. The backward direction works analogously.
	\end{proof}
\subsubsection{Dynamic Programming for \MNTTP on Series-Parallel Networks}
We now present a pseudopolynomial dynamic program that optimally assigns the budget on \sepa networks.
Together with standard rounding techniques, this dynamic program can be extended to a fully polynomial algorithm that achieves the optimal steady-state travel time using a slightly increased budget of $B+\varepsilon\cdot\max\{B, u_0\}$.

\begin{lemma}\label{lem:only need integral solutions}
    Consider a \sepa instance of \MNTTP for which $u_0$ and $\nu$ are integral. Then there is an optimal budget assignment $\xi \in \RR_{\geq 0}^E$ together with a feasible flow $f\in\RR_{\geq0}^E$ in the network $(G,u_0,\nu+\xi)$ such that both $\xi$ and $f$ are integral. (The budget assignment $\xi$ might not use the full budget.)
\end{lemma}
\begin{proof}
    We argue that for each fractional solution (i.e. one that might assign non-integral budget to arcs) there is an integral solution that achieves at most the same steady-state travel time.
    Let $\xi^{\text{frac}} \in \RR_{\geq 0}^E$ be a (fractional) budget assignment and let $f^{\text{frac}}\in \RR_{\geq 0}^E$ be the corresponding steady-state flow. We now want to assign budget so that all flow can travel only on $\text{supp}(f^{\text{frac}})$.
    We construct an auxiliary network $H$ as follows. We set $V(H)=V(G)$. For each arc $e\in\text{supp}(f^{\text{frac}})$, the network contains two arcs with the same start and end points, one with capacity $\nu_e$ and cost $0$, and one with unlimited capacity (equivalently capacity $u_0$) and cost $1$.
    The first arc represents the capacity in $G$ that we can use for free. The second arc represents the additional capacity we can buy with our budget.
    The flow $f^{\text{frac}}$ in $G$ can be turned into a feasible $u_0$-flow in $H$ with cost at most $B$. Hence, there also is an integral $u_0$-flow in $H$ with cost at most $B$. This in turn gives an integral budget assignment (using at most budget $B$) that allows routing all flow integrally using only arcs in $\text{supp}(f^{\text{frac}})$. 
     As $f^{\text{frac}}$ is a feasible flow in $(G,u_0,\nu+\xi^{\text{frac}})$, this integral budget allocation achieves (at most) the same steady-state travel time as the fractional $\xi^{\text{frac}}$ by \cref{lm:support_travel_time}.
\end{proof}
\begin{theorem}\label{lem:SePa_DP}
    The \MNTTP on \sepa networks with integral inflow $u_0$ and integral capacities $\nu$ can be solved optimally in time $\mathcal{O}(|E(G)|\cdot (u_0\cdot B)^2)$.
\end{theorem}
\begin{proof}
    By \Cref{lem:only need integral solutions}, it suffices to consider integral budget assignments (and thus also integral flows).
    Our strategy is to use a dynamic program that determines for a (sub-)network $H$ (with designated source and sink) with inflow $0\leq u\leq u_0$ the minimal travel time one can achieve with a budget of $0\leq b\leq B$. We denote this value by $\TT(H,u,b)$.

    Every \sepa network containing two or more arcs can be composed from smaller \sepa networks using a parallel or a series composition. We fix such a composition and compute $\TT(H,u,b)$ for all subnetworks appearing in this composition and all $0\leq u\leq u_0$, $0\leq b\leq B$.
    We do this in a bottom-up way, starting with the minimal subnetworks (i.e., those containing only one arc). Hence, we need to determine three things: $\TT(H,u,b)$ for such minimal subnetworks, $\TT(H,u,b)$ for $H$ being a series composition of $F$ and $F'$ depending on $\TT(F,\cdot,\cdot)$ and $\TT(F',\cdot,\cdot)$, and finally the same for parallel compositions.
    
    \textbf{Base case: }
    Let $H=(\{s,t\},\{e=(s,t)\})$ be a network containing a single arc $e$ (and its endpoints $s$ and $t$). Let $\tau_e$ and $\nu_e$ be the travel time and capacity of $e$. Then
    \begin{align*}
        \TT(H,u,b)=\begin{cases}
            0 & \text{if } u=0 \\
            \tau_e & \text{if } 0<u\leq\nu_e+b\\
            \infty & \text{else.}
        \end{cases}
    \end{align*}
    
    \textbf{Series composition: }
    Let $H$ arise from the two \sepa networks $F$ and $F'$ using a series composition. Because each flow particle traversing $H$ has to traverse both $F$ and $F'$, the only decision to make is how to distribute the budget between $F$ and $F'$. By \Cref{lm:support_travel_time}, the travel time through $H$ is determined by the sum of the travel times through $F$ and $F'$:
    \begin{align*}
        \TT(H,u,b)=\min_{0\leq b'\leq b} \TT(F,u,b-b')+\TT(F',u,b').
    \end{align*}
    
    \textbf{Parallel composition: }
    Let $H$ arise from the two \sepa networks $F$ and $F'$ using a parallel composition. A flow particle traversing $H$ can either go through $F$ or through $F'$. By \Cref{lm:support_travel_time}, the travel time through $H$ is then determined by the maximum travel time through $F$ and through $F'$. In a steady state, the flow is distributed between $F$ and $F'$ such that this maximum is minimized. Hence, we obtain the following recursion:
    \begin{align*}
        \TT(H,u,b) = \min_{\substack{0 \leq u' \leq u\\0 \leq b' \leq b}} \max\{\TT(F, u-u', b-b'),\ \TT(F',u', b')\}.
    \end{align*}
    There are at most $2|E(G)|$ subnetworks in our composition of $G$. Hence, we compute $\mathcal{O}(|E(G)|\cdot u_0\cdot B)$ many entries for our dynamic program. Each entry can be computed in time $\mathcal{O}(u_0\cdot B)$ using the recursion above. This results in a total run time in $\mathcal{O}(|E(G)|\cdot (u_0\cdot B)^2)$. The optimal budget assignment can be extracted by backtracking how $\TT(G,u_0,B)$ was computed.
\end{proof}
\begin{theorem}
    Fix $\varepsilon>0$ small and let $\mathcal I=\instance$ be an \MNTTP instance with $G$ being a \sepa network.
There is an $\varepsilon$-budget augmentation algorithm for $\mathcal I$ with runtime in $\mathcal{O}(\frac{|E|^5}{\varepsilon^4})$.
\end{theorem}
\begin{proof}
    Without loss of generality, $\nu_e\leq u_0$ for all $e\in E$.
    We round the instance such that \Cref{lem:SePa_DP} can be applied using only a polynomial runtime (in the size of $G$).
    Let $\delta \coloneqq \frac{\varepsilon}{2|E(G)|}\cdot\max\{B,u_0\}$. To construct our rounded instance, define $\tilde\nu\coloneqq \lceil\nu\rceil_{\delta}$, $\tilde u_0\coloneqq\lfloor u_0\rfloor_{\delta}$, and $\tilde B\coloneqq\lfloor B\rfloor_{\delta}$
    where $\lceil y\rceil_\delta\coloneqq\min\{x\in\delta\mathbb{Z}\ :\ x\geq y\}$ and $\lfloor y\rfloor_\delta\coloneqq\max\{x\in\delta\mathbb{Z}\ :\ x\leq y\}$.

    By \Cref{lem:SePa_DP}, the instance $(G,\tilde u_0,\tilde\nu,\tau,\tilde B)$ can be solved optimally in time $\mathcal{O}(|E(G)|\cdot(u_0/\delta\,\cdot\,B/\delta)^2)=\mathcal{O}(|E(G)|^5/\varepsilon^4)$.
    Let $\TT$ and $\tilde{\TT}$ be the optimal travel time for instances $\instance$ and $(G,\tilde u_0,\tilde\nu,\tau,\tilde B)$.
    We claim two things. First, $\tilde{\TT}\leq\TT$. Second, given an (optimal) solution to the instance $(G,\tilde u_0,\tilde\nu,\tau,\tilde B)$, we can turn it into a solution of the instance $(G,u_0,\nu,\tau,B+\varepsilon\cdot\max\{B,u_0\})$ with steady-state travel time at most \TTT. If $\tilde u_0 >0$, the bound is even $\tilde \TT$.

    For the first claim, \cref{prop:monotonicty_in_s-p_networks} implies that increasing $\nu$ cannot increase the optimal steady-state travel time. The same holds for decreasing $u_0$, as scaling up $u_0$ by a factor $C$ is equivalent to scaling down $\nu$ by $C$. 
    Furthermore, by \Cref{lem:only need integral solutions}, if $u_0$ and all capacities are multiples of $\delta$, there is an optimal solution assigning the budget in multiples of $\delta$. Hence, rounding down $B$ as well does not affect the optimal travel time, yielding the first claim.

    For the second claim, we first assume that $\tilde u_0>0$. Consider an optimal budget assignment $\tilde\xi \in \RR_{\geq 0}^E$ for the instance $(G,\tilde u_0,\tilde\nu,\tau,\tilde B)$. By \Cref{lm:support_travel_time}, there exists a flow $\tilde f\in \RR_{\geq 0}^E$ in $G$ of value $\tilde u_0$ with $\tilde f\leq\tilde\nu+\tilde\xi$ such that the longest $s$-$t$ path in the support of $\tilde f$ has length (at most) $\tilde \TT$.
    By adding flow of value $u_0-\tilde u_0$ to some $s$-$t$ path in the support of $\tilde f$ (which exists by $\tilde u_0>0$) we obtain a flow $f$ with
    \begin{equation*}
         f\leq(u_0-\tilde u_0)\cdot\mathbf{1}+\tilde f\leq \delta\cdot\mathbf{1}+\tilde f\leq\delta\cdot\mathbf{1}+\tilde\nu+\tilde\xi\leq\nu+2\delta\cdot\mathbf{1}+\tilde\xi\,.
    \end{equation*}
    Hence, choosing $\xi=2\delta\cdot\mathbf{1}+\tilde\xi$ suffices to ensure that there is a flow $f$ of value $u_0$ with $f\leq\nu+\xi$, whose support is no larger than the support of $\tilde f$. By \Cref{lm:support_travel_time}, this implies that the steady-state travel time for budget assignment $\xi$ is bounded by $\tilde\TT$.
    The total sum of the $\xi$ values differs from the total sum of the $\tilde\xi$ values by $2|E(G)|\cdot\delta\leq \varepsilon\cdot\max\{B,u_0\}$.

    If $\tilde u_0=0$, then $u_0<\delta$. Hence, in this case, we can achieve the best possible travel time by assigning $\delta$ units of budget to each arc of some shortest $s$-$t$ path. This uses a total budget of at most $|E(G)|\cdot\delta<\varepsilon\cdot\max\{B,u_0\}$.
\end{proof}
\begin{remark}
     An interesting related problem is reducing transit times with a budget.
     It can easily be seen that this problem is already NP-hard on a single bundle by a reduction from \textsc{Knapsack}.
     On \sepa networks, this problem is solvable with a similar dynamic program to the capacity version above with the caveat that for transit times fractional changes can have an impact on \TTT. Consequently, a corresponding hyperplane arrangement would have regions in which \TTT is piecewise linear. Combining the dynamic programs for capacities and transit times, one can optimize both simultaneously.
\end{remark}

\subsection{NP-Hardness and Inapproximability of \MNTTP on General Networks}
\label{sec:general_graphs}

\begin{figure}[]
    \centering

 \scalebox{0.9}{
\begin{tikzpicture}[
    yscale=0.6, xscale=0.85,
    vertex/.style={circle, fill=black, inner sep=1.6pt},
    >=stealth,
    font=\small
]
 \tikzset{
  ellipsoid/.style={
    draw, ellipse,
    minimum width=80pt,
    minimum height=30pt,
    inner sep=0pt
  }
}

\draw[thick, rounded corners=12pt, fill=gray!10]
    (-0.5, -3.8) rectangle (5.0, 3.8);
\node at (2.3, 0.25) {\large $G$};

\node[shape=circle, scale=\scl, draw=black] (S) at (-6,  0) {$s$};
\node[shape=circle, scale=\scl, draw=black] (T) at (11,  0) {$t$};

\node[shape=circle, scale=\scl, draw=black] (S1) at (0.2,  3)  {$s_1$};
\node[shape=circle, scale=\scl, draw=black] (S2) at (0.2,  1)  {$s_2$};
\node[shape=circle, scale=\scl, draw=black] (Sk) at (0.2, -3)  {$s_k$};
 
\node[shape=circle, scale=\scl, draw=black] (T1) at (4.2,  3)  {$t_1$};
\node[shape=circle, scale=\scl, draw=black] (T2) at (4.2,  1)  {$t_2$};
\node[shape=circle, scale=\scl, draw=black] (Tk) at (4.2, -3)  {$t_k$};

\node at (0.2, -1) {$\vdots$};
\node at (4.2, -1) {$\vdots$};

\node[ellipsoid] (leftTop)    at (-2.5,  3) {};
\node[ellipsoid] (leftMiddle) at (-2.5,  1) {};
\node[ellipsoid] (leftBottom) at (-2.5, -3) {};
\node at (-2.5, -1) {$\vdots$};
 
\foreach \y [count=\row] in {3,1,-3} {
    \node[vertex] (a\row) at (-3.5,\y) {};
    \node[vertex] (b\row) at (-3.0,\y) {};
    \node         (e\row) at (-2.25,\y) {$\cdots$};
    \node[vertex] (c\row) at (-1.5,\y) {};
    \draw[->] (a\row) -- (b\row);
    \draw[->] (b\row) -- (e\row);
    \draw[->] (e\row) -- (c\row);
}

\node[ellipsoid] (rightTop)    at (7,  3) {};
\node[ellipsoid] (rightMiddle) at (7,  1) {};
\node[ellipsoid] (rightBottom) at (7, -3) {};
\node at (7, -1) {$\vdots$};
 
\foreach \y [count=\row] in {3,1,-3} {
    \node[vertex] (x\row) at (6,\y) {};
    \node[vertex] (y\row) at (6.5,\y) {};
    \node         (f\row) at (7.25,\y) {$\cdots$};
    \node[vertex] (z\row) at (8,\y) {};
    \draw[->] (x\row) -- (y\row);
    \draw[->] (y\row) -- (f\row);
    \draw[->] (f\row) -- (z\row);
}
 
\path (S) edge[->, bend left=20]
    node[midway, sloped, below, yshift=0.5pt, rotate = -5] {$(1,0)$}
    (a1);
\path (S) edge[->]
    node[midway, sloped, below, yshift=0.5pt, rotate = -4] {$(1,n)$}
    (a2);
\path (S) edge[->, bend right=20]
    node[midway, sloped, above, yshift=0.5pt, rotate=9] {$(1,(k{-}1)n)$}
    (a3);
 
\path (z1) edge[->, bend left=20]
    node[midway, sloped, below, yshift=0.5pt, rotate=10] {$(1,(k{-}1)n)$}
    (T);
\path (z2) edge[->]
    node[midway, sloped, below, xshift=2.5pt, rotate=5] {$(1,(k{-}2)n)$}
    (T);
\path (z3) edge[->, bend right=20]
    node[midway, sloped, above, yshift=0.5pt, rotate=-10] {$(1,0)$}
    (T);

\path (c1) edge[-] (leftTop.east);
\path (c2) edge[-] (leftMiddle.east);
\path (c3) edge[-] (leftBottom.east);
 
\path (rightTop.west)    edge[->] (x1);
\path (rightMiddle.west) edge[->] (x2);
\path (rightBottom.west) edge[->] (x3);

\path (leftTop.east)    edge[->] (S1);
\path (leftMiddle.east) edge[->] (S2);
\path (leftBottom.east) edge[->] (Sk);
 
\path (T1) edge[-] (rightTop.west);
\path (T2) edge[-] (rightMiddle.west);
\path (Tk) edge[-] (rightBottom.west);

\path (S) edge[->, bend left=90, looseness=1.1]
    node[midway, sloped, below, yshift=-3pt] {$(k+1,\,T_{\max})$}
    (T);

\draw [decorate,decoration={brace,amplitude=5pt,mirror,raise=4ex}]
  (-4.2,-3) -- (-0.8,-3) node[midway,yshift=-3em]{$M$ nodes each};
\draw [decorate,decoration={brace,amplitude=5pt,mirror,raise=4ex}]
  (5.3,-3) -- (8.7,-3) node[midway,yshift=-3em]{$M$ nodes each};
\end{tikzpicture}}
    \caption{The constructed \MNTTP instance for the reduction in the proofs of \cref{thm:strongnphard,thm:inapprox}.
    \label{fig:strongly_reduction}
    }
\end{figure}
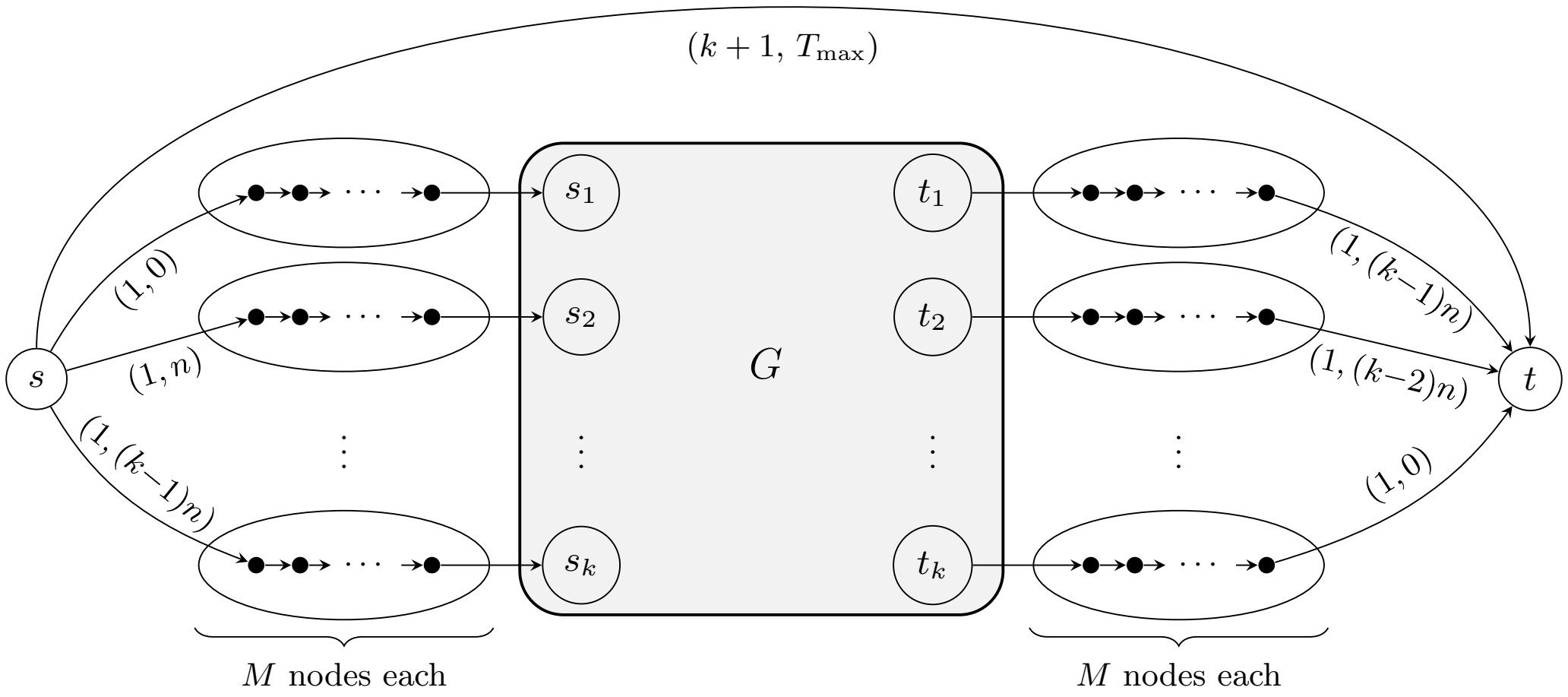

On general networks, \MNTTP is strongly NP-hard, by a reduction from \textsc{$k$-Disjoint Paths}. In \textsc{$k$-Disjoint Paths} a directed graph $G$ with $k$ source-sink pairs $(s_i,t_i)_{i\in [k]}$ is given. The task is to decide whether there exist vertex- or edge-disjoint $s_i$-$t_i$-paths in $G$, one for each pair of terminals. If $k$ is part of the input, this problem is NP-hard \cite{fortune1980directed} even on directed acyclic graphs (DAGs) \cite{even1975complexity}. The vertex and edge disjoint versions of this problem are well-known to be equivalent (see e.g. \cite{Schrijver03CombiOptim}); we consider the vertex disjoint version of \textsc{$k$-Disjoint Paths}. We assume without loss of generality that there exists an $s_i$-$t_i$-path in the input network for every $i\in [k]$ and that all terminals are distinct; otherwise, we cannot hope for vertex-disjointness. 

The idea is the following: Given an instance of \textsc{$k$-Disjoint Paths} with $n$ nodes, we introduce a super-source $s$ with inflow rate $u_0=k$ and a super-sink $t$ together with an $(s,t)$-arc with capacity $k+1$ and travel time $T_{max}$ to provide a baseline travel time. We show that any algorithm for \MNTTP with budget at most $M$ can decide \textsc{$k$-Disjoint Paths} based on whether or not it can obtain $\TT<T_{\max}$. To this end, we connect $s$ to each $s_i$ and $t_i$ to $t$ via chains of $M+1$ arcs. Each chain has capacity 1, this forces the algorithm to spend its budget inside $G$, as the budget is insufficient to increase the capacity of a long chain to support an additional unit of flow.  Moreover, the travel times on the chains are chosen to enforce disjoint paths; if the paths of $i< j$ cross at $v$, particles first choose the shorter $s_i$-$v$-$t_j$-path on which queues grow until $s_j$-$v$-$t_i$ becomes active, yielding a travel time of at least $T_{max}$ by \cref{obs:longest_path_lower_bound}.  

\textbf{Construction of the \MNTTP instance:} Given an instance $(G,(s_i,t_i)_{i\in [k]})$ of \textsc{$k$-Disjoint Paths}, we create an instance of \MNTTP with $\tilde G = (\tilde V, \tilde E)$ as follows. Let $n\coloneqq |V|$ and define the node set to be $\tilde V \coloneqq V \cup \{s,t\} \cup \bigcup_{i=1}^k\{s_i^1, \dots, s_i^{M}, t_i^1, \dots, t_i^{M}\}$ for some $M\in \mathbb N$ with $M\geq n$. The arc set $\tilde E$ consists of the arc set $E$ together with an arc $(s,t)$ as well as arcs $E^{s,t}\coloneqq \{(s, s_i^1), (s_i^j, s_i^{j+1}), (s_i^{M},s_i), (t_i,t_i^1),(t_i^j, t_i^{j+1}), (t_i^{M},t) \mid i\in [k], j\in [M-1]\}$. Thus, both the super-source $s$ and the super-sink $t$ are connected to each of the $s_i$, $t_i$ respectively by a path containing $M$ internal vertices. 
We set as inflow $u_0=k$ and set the capacities to $\nu_{st}=k+1$, $\nu_e=0$ for $e\in E$ and $\nu_{\tilde e}=1$ for $\tilde e\in  E^{s,t}$. Moreover, we set the transit times to be $\tau_{st}=T_{\max}=2M+(n-2)k+2$ and $\tau_{ss_i^1}=(i-1)n,\ \tau_{t_i^{M}t}=(k-i)n$ for $i\in [k]$, all other transit times are set to 1. We set the budget to be $B\coloneqq n-k$. Note that if $G$ is a DAG, then so is $\tilde G$. The construction is illustrated in \cref{fig:strongly_reduction}.
        
\begin{observation}\label{claim:too_long_path}
    Consider a budget allocation yielding $\TT<T_{\max}$. Then no active path in the steady state of a \NFOT can connect $s_i$ and $t_j$ with $i>j$.
\end{observation}

\begin{proof}
    An $s$-$t$ path $\tilde P$ using such an $s_i$-$t_j$-path has length $\sum_{e\in \tilde P} \tau_e \geq 2 M+kn+1 > 2M+(n-2)k+2 = T_{\max}$. As it is active, the path length yields a lower bound on \TTT.  
\end{proof}
        
The core ingredient of the hardness and inapproximability proofs is the following lemma. Whenever we invoke previous lemmata, we only consider the network without 0-capacity arcs and as usual apply an \perturb after the budget allocation so that $\nu\notin \mathbb H_G$.
\begin{lemma}\label{lm:yes_instance}
    If there exists an allocation of a budget of at most $M$ achieving $\TT<T_{\max}$, then $(G,(s_i,t_i)_{i\in [k]})$ is a \textsc{Yes}-instance of \textsc{$k$-Disjoint Paths}.
\end{lemma}
\begin{proof}
    Assume that we have a budget allocation yielding $\TT < T_{\max}$ and consider the steady-state flow $f$ of some \NFOT. Clearly, $f_{st}=0$. We show that then the support of $f$ must contain disjoint $s_i$-$t_i$-paths for $i\in [k]$, and thus we have a \textsc{Yes}-instance.
    We can assume without loss of generality that if fractional budget is allocated to an $s$-$s_i$-path or a $t_i$-$t$-path, then the same amount of budget $b_{s_i}$, respectively $b_{t_i}$, is allocated to each arc on the path. Let $b_s\coloneqq \sum_{i\in [k]} b_{s_i}$ be the cumulative capacity increase on the set of $s$-$s_i$-paths and $b_t$ that on the set of $t_i$-$t$-branches. The total budget spent can be lower bounded by $(M+1)\cdot(b_s+b_t)$. Hence, $b_s+b_t<1$.
    Consider any subset $X$ of $\{s_i\}_{i\in[k]}$ and let $f(X)$ denote the total through flow of the vertices in $X$. If $f(X)$ is less than $|X|$, the difference $|X|-f(X)$ has to travel through some other $s$-$s_i$-paths, but all paths from $s$ to $\{s_i\}_{i\in[k]}\setminus X$ can only support an additional flow of $b_s$. Thus, $f(X) \geq |X| -b_s$. Now let $\mathcal P$ denote the set of $s$-$t$ paths chosen as strategy by the flow particles in the \NFOT steady state such that each $P\in \mathcal P$ carries positive flow. Let $N(X)$ be the subset of $\{t_i\}_{i\in[k]}$ reachable from $X$ via paths in $\mathcal P$. Observe that $|X|-b_s \leq f(X) \leq f(N(X)) \leq |N(X)|+b_t$. Thus, $|X|-(b_s+b_t) \leq  |N(X)|$, but $b_s+b_t<1$, and we can choose the \perturb small enough for the resulting inequality to remain strict.
    It follows by Hall's Theorem that there must be a perfect matching consisting of $s_i$-$t_j$-paths in the support of $f$. But the support of $f$ cannot contain any $s_i$-$t_j$ paths for $i>j$ by \cref{claim:too_long_path}. Thus, the perfect matching also cannot contain any $s_i$-$t_j$ paths for $i<j$ and thus consists entirely of $s_i$-$t_i$-paths. Moreover, we cannot have a \textsc{No}-instance, as in that case for any choice of $k$ paths there exist $i,j\in [k]$ such that the $s_i$-$t_i$- and $s_j$-$t_j$-paths share a vertex $v$. Without loss of generality, $i>j$. But as all four segments $s_i$-$v$, $s_j$-$v$, $v$-$t_i$ and $v$-$t_j$ are active since they are flow-carrying, so is the $s$-$t$ path via $s_i,v$ and $t_j$. But then $\TT \geq T_{\max}$ by \cref{claim:too_long_path}, a contradiction.
\end{proof}

\begin{restatable}{theorem}{thmstrongnphard}
\label{thm:strongnphard}
    The \MNTTP on general networks is strongly NP-hard, even on DAGs.
\end{restatable}
    
We directly prove the stronger result that \MNTTP is hard to approximate, even on DAGs. 
\begin{theorem}\label{thm:inapprox}
    Unless $\text{P} = \text{NP}$, for every fixed $\varepsilon \in (0,1)$, there is no better-than-$(N^{1 - \varepsilon})$-budget-augmentation algorithm for the \MNTTP, even on DAGs. Here $N$ denotes the number of nodes of the \MNTTP instance.
\end{theorem}

\begin{proof}
    Suppose there is a budget augmentation algorithm with augmentation ratio less than $N^{1-\varepsilon}$ for some $\varepsilon \in (0,1)$. Choose some integer $c > \tfrac{2}{\varepsilon}-1$ and let $M = n^c$. Thus, $N= n + 2kM + 2\in \mathcal{O}(kn^c)\subseteq \mathcal O(n^{c+1})$. Let $B = n-k$, then as $u_0=k\leq \tfrac n2$ an $N^{1-\varepsilon}$ budget augmentation algorithm spends at most $(1+N^{1-\varepsilon})B \in \mathcal O(n^{(c+1)(1-\varepsilon)+1})\subseteq o(M)$. This decides \textsc{$k$-Disjoint-Paths} in polynomial time on sufficiently large instances. On a \textsc{Yes}-instance, the algorithm obtains a solution with $\TT < T_{\max}$ which is only possible by finding disjoint $s_i$-$t_i$-paths for $i\in[k]$ due to the augmented budget being bounded above by $M$. By \cref{lm:yes_instance}, on a \textsc{No}-instance, no solution with $\TT < T_{\max}$ can be found. Therefore, no such budget augmentation algorithm can exist unless $\text{P}=\text{NP}$.
\end{proof}

\newpage
\bibliographystyle{plainurl}
\bibliography{references}

@article{yang1998models,
  title={Models and algorithms for road network design: a review and some new developments},
  author={Yang, Hai and H. Bell, Michael G},
  journal={Transport Reviews},
  volume={18},
  number={3},
  pages={257--278},
  year={1998},
  publisher={Taylor \& Francis}
}

@inproceedings{even1975complexity,
  title={On the complexity of time table and multi-commodity flow problems},
  author={Even, Shimon and Itai, Alon and Shamir, Adi},
  booktitle={16th annual symposium on foundations of computer science (sfcs 1975)},
  pages={184--193},
  year={1975},
  organization={IEEE}
}

@article{fortune1980directed,
  title={The directed subgraph homeomorphism problem},
  author={Fortune, Steven and Hopcroft, John and Wyllie, James},
  journal={Theoretical Computer Science},
  volume={10},
  number={2},
  pages={111--121},
  year={1980},
  publisher={Elsevier}
}

@article{braess1968paradoxon,
  title={{\"U}ber ein {Paradoxon} aus der {Verkehrsplanung}},
  author={Braess, Dietrich},
  journal={Unternehmensforschung},
  volume={12},
  number={1},
  pages={258--268},
  year={1968},
  publisher={Springer}
}

@inproceedings{correa2019price,
  title={On the price of anarchy for flows over time},
  author={Correa, Jos{\'e} and Cristi, Andr{\'e}s and Oosterwijk, Tim},
  booktitle={Proceedings of the 2019 ACM Conference on Economics and Computation},
  pages={559--577},
  year={2019}
}

@article{koch2011nash,
  title={Nash equilibria and the price of anarchy for flows over time},
  author={Koch, Ronald and Skutella, Martin},
  journal={Theory of Computing Systems},
  volume={49},
  number={1},
  pages={71--97},
  year={2011},
  publisher={Springer}
}

@book{koch2012routing,
  title={Routing games over time},
  author={Koch, Ronald},
  year={2012},
  publisher={Technische Universität Berlin (Germany)}
}

@article{vickrey1969congestion,
  title={Congestion theory and transport investment},
  author={Vickrey, William S.},
  journal={The American Economic Review},
  volume={59},
  number={2},
  pages={251--260},
  year={1969},
  publisher={JSTOR}
}

@article{chen2026network,
  title={Network Topology That Excludes {Braess's} Paradox and Maintains Monotonicity in Flows Over Time},
  author={Chen, Xujin and Deng, Xiyuan and Wang, Changjun},
  journal={arXiv preprint arXiv:2609.32738},
  year={2026}
}

@unpublished{Correa2026Counterexample,
  author = {Correa, Jos{\'e}},
  title  = {A Counterexample to the Monotonicity Conjecture
            for Dynamic Equilibria in Flows over Time},
  note   = {October 4, 2026},
  url    = {https://www.dii.uchile.cl/~jcorrea/papers/Manuscripts/2022CCOcounterexample.pdf}
}

@article{macko2013braess,
  title={Braess’s paradox for flows over time},
  author={Macko, Martin and Larson, Kate and Steskal, L'ubo{\v{s}}},
  journal={Theory of Computing Systems},
  volume={53},
  number={1},
  pages={86--106},
  year={2013},
  publisher={Springer}
}

@inproceedings{chen2015excluding,
  title={Excluding {Braess’s} paradox in nonatomic selfish routing},
  author={Chen, Xujin and Diao, Zhuo and Hu, Xiaodong},
  booktitle={International Symposium on Algorithmic Game Theory},
  pages={219--230},
  year={2015},
  organization={Springer}
}

@article{milchtaich2006network,
  title={Network topology and the efficiency of equilibrium},
  author={Milchtaich, Igal},
  journal={Games and Economic Behavior},
  volume={57},
  number={2},
  pages={321--346},
  year={2006},
  publisher={Elsevier}
}

@article{cominetti2015dynamic,
  title={Dynamic equilibria in fluid queueing networks},
  author={Cominetti, Roberto and Correa, Jos{\'e} and Larr{\'e}, Omar},
  journal={Operations Research},
  volume={63},
  number={1},
  pages={21--34},
  year={2015},
  publisher={INFORMS}
}

@article{cominetti2022long,
  title={Long-term behavior of dynamic equilibria in fluid queuing networks},
  author={Cominetti, Roberto and Correa, Jos{\'e} and Olver, Neil},
  journal={Operations Research},
  volume={70},
  number={1},
  pages={516--526},
  year={2022},
  publisher={INFORMS}
}

@inproceedings{olver2022continuity,
  title={Continuity, uniqueness and long-term behavior of {Nash} flows over time},
  author={Olver, Neil and Sering, Leon and Koch, Laura Vargas},
  booktitle={2021 IEEE 62nd Annual Symposium on Foundations of Computer Science (FOCS)},
  pages={851--860},
  year={2022},
  organization={IEEE}
}

@article{bhaskar2015stackelberg,
  title={A {Stackelberg} strategy for routing flow over time},
  author={Bhaskar, Umang and Fleischer, Lisa and Anshelevich, Elliot},
  journal={Games and Economic Behavior},
  volume={92},
  pages={232--247},
  year={2015},
  publisher={Elsevier}
}

@inproceedings{bhaskar2014network,
  title={Network improvement for equilibrium routing},
  author={Bhaskar, Umang and Ligett, Katrina and Schulman, Leonard J.},
  booktitle={International Conference on Integer Programming and Combinatorial Optimization},
  pages={138--149},
  year={2014},
  organization={Springer}
}

@article{gairing2017complexity,
  title={Complexity and approximation of the continuous network design problem},
  author={Gairing, Martin and Harks, Tobias and Klimm, Max},
  journal={SIAM Journal on Optimization},
  volume={27},
  number={3},
  pages={1554--1582},
  year={2017},
  publisher={SIAM}
}

@book{Schrijver03CombiOptim,
  title = {Combinatorial {{Optimization}}},
  author = {Schrijver, Alexander},
  date = {2003},
  year = {2003},
  series = {Algorithms and {{Combinatorics}}},
  edition = {1st},
  volume = {24},
  publisher = {Springer Nature},
  isbn = {978-3-540-44389-6}
}
\newpage
\appendix
\section{Appendix}
\subsection{Omitted Proofs}

\multiplicativeeps*
\begin{proof}  
    Note that perturbing $u_0$ by a factor of $(1-\delta)$ for some $\delta \in \mathbb{R}$ is equivalent to perturbing $\nu$ by a factor of $(1+\eta)$ for $\eta = \delta/(1-\delta)$. This holds as the objective of \lpref{D} for $(G,(1-\delta)u_0,\nu,\tau)$ and $(G,u_0,(1+\eta)\nu,\tau)$ differs only by a constant multiplicative factor $(1-\delta)$ and the feasible region remains unaffected.

    Let Sol$(D_{u_0})$ denote the solution polyhedron of (\ref{D}) with inflow rate $u_0$.
    First, we show that there exists $\gamma\in (0,1)$ such that for all perturbations $\delta \in  [-\gamma,\gamma]$ of the inflow rate $u_0$ we have Sol$(D_{(1-\delta)u_0}) \subseteq \text{Sol}(D_{u_0})$. Let $\text{obj}^i(\delta)=(1-\delta)u_0d_t^i - \sum_e\nu_eq_e^i$ be the objective value at a vertex $(d_i,q_i)$ of the feasible region as we perturb the inflow and note that $\text{obj}^i(\delta)$ is continuous in $\delta$. For a suboptimal $(d^i,q^i)\not\in \text{Sol}(D_{u_0})$ we have $\text{obj}^i(0)<u_0{d_*}_t-\sum_e \nu_e{q_*}_e$. Thus, continuity implies that there exists $\gamma \in (0,1)$ such that $(d^i,q^i)$ remains suboptimal for all $(1-\delta)u_0$ with $\delta \in [-\gamma,\gamma]$. 
    
    Second, note that for $\delta > 0$, the reduction in the objective $(1-\delta)u_0d_t -\sum_{e \in E}\nu_e q_e$ compared to $\delta = 0$ is smallest for those solutions $(d,q) \in \text{Sol}(D_{u_0})$ with minimal $d_t$-value. 
    For $\delta < 0$, the increase in objective value is largest for those solutions with maximal $d_t$-value. 
     As Sol$(D_{(1-\delta)u_0}) \subseteq \text{Sol}(D_{u_0})$, the statement follows.
\end{proof}

\thmstrongnphard*
    \begin{proof}
        We choose $M=n$ and claim that an instance of \textsc{$k$-Disjoint Paths} is a \textsc{Yes}-instance if and only if $\TT<T_{\max}$.
        Given a \textsc{Yes}-instance, we can allocate one unit of capacity to each arc in the support of the $k$ vertex-disjoint paths in the solution at cost at most 
        $n-k$, the bound being tight if all nodes of $G$ are visited by the $k$ paths. Then, the $k$ internally vertex-disjoint $s-t$ paths $P_i$ in $\tilde G$ all satisfy $\sum_{e\in P_i} \tau_e \leq 2M+(k-1)n+n-2k+1=2M+(n-2)k+1$. 
        The resulting network restricted to arcs with positive capacity is \sepa and the capacity of $\tilde G - st$ after an \perturb suffices to support the \NFOT, thus,
        $\TT \le 2M+(n-2)k+1 < T_{\max}$ by \cref{lm:support_travel_time}. 
         \cref{lm:yes_instance} gives the reverse direction.
    \end{proof}

\subsection{Additional Structural Result}
Recall that given a network with data $(G,u_0,\tau)$, we denote by $\steadytime(\nu)$ the time to reach steady state and by $\TT(\nu)$ the steady-state travel time of the \NFOT in \network. 
We show that if a hyperplane divides two regions with different \TTT, the time to reach steady state diverges to $\infty$ if the hyperplane is approached from at least one side.

\begin{proposition}
    Consider two cells $C_1$ and $C_2$ of the hyperplane arrangement $\mathbb{H}_G$ with non-empty intersection. Let $\nu_i$ be a point in the interior $\intr(C_i)$ for $i\in \{1,2\}$ and let $\bar \nu \in C_1 \cap C_2$. If $\TT(\nu_1)\neq \TT(\nu_2)$ then for every $i\in\{1,2\}$ such that $\TT(\nu_i)\neq \TT(\bar \nu)$ (of which there is at least one),
    \begin{equation*}
        \lim_{\substack{\nu \to \bar\nu\\ \nu\in \intr(C_i)}}\steadytime(\nu) = \infty.
    \end{equation*}
\end{proposition}

\begin{proof}
    As $\TT(\nu_1)\neq \TT(\nu_2)$, we must have $\TT(\nu_i)\neq \TT(\bar \nu)$ for at least one $i\in[2]$, without loss of generality for $\nu_1$. Now assume for a contradiction that $\steadytime(\nu)\not\to \infty$ as $\nu \in C_1$ approaches $\bar \nu$. Then there exists a bound $M>0$ as well as a sequence $(\nu_j)_{j\in \mathbb N}$ converging to $\bar \nu$ such that $\nu_j \in \intr(C_1)$ and $\steadytime(\nu_j) < M$ for all $j$. Let $N = \max\{\steadytime(\bar \nu),M\}$ be a time at which in all networks $(G,u_0,\nu_j,\tau)$ and $(G,u_0,\bar\nu,\tau)$ the steady state is reached. As the steady-state travel time of all particles in a network is exactly \TTT, by \cref{thm:d_t_constant_in_region} we have 
    \begin{align}\label{eq:steady-state_times_diverge}
        \ell^{\nu_j}_t(N)-N = \TT(\nu_1) \text{ for all } j \quad \text{ and } \quad \ell_t^{\bar \nu}(N)-N = \TT(\bar \nu),
    \end{align}
    where $\ell^\nu_t(\sourceTime)$ is the earliest arrival time at $t$ of a particle departing the source at time $\sourceTime$ and traveling through \network.
    Since $(\nu_j)_{j\in \mathbb N}$ converges to $\bar \nu$, $\TT(\nu_1)\neq \TT(\bar \nu)$ together with \eqref{eq:steady-state_times_diverge} directly contradicts the continuity of $\ell$ with respect to $\nu$ at time $N$ (Theorem 3.6 of \cite{olver2022continuity}). 
\end{proof}

\begin{example}
    Consider again the example from \cref{fig:unnecessary queue}, with an additional $wt$ arc $e_6$ with data $(\mu,10)$ and consider \TTT as the capacity $\nu_{e_5}$ varies over $[12-\varepsilon,12+\varepsilon]$ for some sufficiently small $\varepsilon>0$. For the minimum cut condition to be satisfied, we require $\mu > \varepsilon$. The steady-state travel time from $s$ to $w$ remains unaffected. We thus only consider the steady-state travel time $\ell_t-\ell_w$ from $w$ to $t$. For $\nu_{e_5}<12$, $\ell_t-\ell_w=10$ as a queue builds up on $e_5$ until arc $e_6$ becomes active. For $\nu_{e_5}=12$, $\ell_t-\ell_w=\tfrac 3 {70}$, the queueing time caused by the ``unnecessary'' queue on $e_5$. Lastly, for $\nu_{e_5} > 12$, $\ell_t-\ell_w=0$ as the ``unnecessary'' queue can deplete and arc $e_6$ never becomes active. 
    Thus, at $\nu_{e_5}=12$, $\TT = \ell_t-\ell_s$ lies strictly between the minimum and the maximum $d_t$ value of any optimal dual solution. This also implies that as $\varepsilon\to 0$, the time to reach steady state in the networks with $\nu_{e_5}=12-\varepsilon$ and $\nu_{e_5}=12+\varepsilon$ both diverge to infinity.
\end{example}
\end{document}